\documentclass[prb,twocolumn,superscriptaddress,citeautoscript,showpacs,amsart]{revtex4}

\usepackage{graphicx}
\usepackage{multirow}
\usepackage{color}
\usepackage{bm}
\usepackage{times}
\usepackage{amsmath,bm,amsfonts}
\usepackage{dcolumn}
\usepackage{graphicx}
\usepackage{latexsym}

\usepackage{ulem} 
\usepackage{braket}

\usepackage{cancel}

\begin{document}
\title{
From Berry curvature to quantum metric: a new era of quantum geometry metrology for Bloch electrons in solids
}

\author{Bohm-Jung \surname{Yang}}
\email{bjyang@snu.ac.kr}
\affiliation{Department of Physics and Astronomy, Seoul National University, Seoul 08826, Korea}

\affiliation{Center for Theoretical Physics (CTP), Seoul National University, Seoul 08826, Korea}

\affiliation{Institute of Applied Physics, Seoul National University, Seoul 08826, Korea}

\date{\today}

\begin{abstract}
For decades, “geometry” in band theory has largely meant Berry phase and Berry curvature—quantities that reshape semiclassical dynamics and underpin modern topological matter. Yet the full geometric content of a Bloch band is richer and encoded in the quantum geometric tensor (QGT), whose imaginary part is the Berry curvature and whose real part is the quantum metric. Here, we briefly review the recent progress in direct experimental access to the QGT in real crystalline solids using the polarization- and spin-resolved angle-resolved photoemission spectroscopy (ARPES). The extraction of the QGT in momentum space was successfully addressed by two different approaches: One is by introducing quasi-QGT that faithfully represents the QGT and is directly measurable by ARPES. The other is through pseudospin tomography in a material with simple low energy band structure, which successfully retrieved all matrix components of the quantum metric. We discuss the physical meaning of these two recent progresses, their implication/limitation, and open directions.
\end{abstract}

\maketitle








The geometric structure of Bloch wavefunctions has emerged as a unifying principle in modern condensed matter physics~\cite{old1, old2, review1, review2, review3}. Beyond energy dispersions, the internal geometry of electronic states in momentum space governs a growing class of physical phenomena, ranging from topological transport~\cite{AHE} to geometric superconductivity~\cite{Torma1, Bockrath,Torma2, Bernevig, Torma3}, anomalous Landau levels~\cite{LL1, LL2, LL3}, and nonlinear responses~\cite{BCdipole, JAhn, QMdipole1, QMdipole2}. This geometry is compactly encoded in the quantum geometric tensor (QGT), whose imaginary part gives the Berry curvature and whose real part defines the quantum metric, a measure of the distance between neighboring quantum states in momentum space. While Berry curvature has become experimentally tangible over the past decade~\cite{CD1, CD2, CD3, CD4}, the quantum metric has long remained elusive in real solids. Two recent breakthroughs~\cite{Yang1, Yang2} now decisively change this situation, demonstrating that quantum geometry can be directly measured in crystalline materials rather than inferred indirectly.

The first work by Kang et al.~\cite{Yang1} develops a general framework to reconstruct the full QGT, including both the quantum metric and Berry curvature, by combining band-curvature analysis and spin-resolved circular dichroism angle-resolved photoemission spectroscopy (ARPES), with a demonstration in the kagome metal CoSn. The second work by Kim et al.~\cite{Yang2} reports a direct measurement of the full quantum metric tensor for Bloch electrons in a real material (black phosphorus) using polarization-dependent ARPES. Together, they effectively convert ``quantum geometry'' from a derived theoretical layer into an experimentally chartable set of band- and momentum-resolved fields, comparable in spirit to how ARPES made electronic dispersion a directly measurable object.

{\it Why the quantum metric was hard and why ARPES is the right lever.}---
The Berry curvature has long dominated discussions of band geometry because of its clear physical manifestations: anomalous velocities, quantized Hall responses, and topological invariants~\cite{Niu}. By contrast, the quantum metric describes the spread and deformation of wavefunctions in Hilbert space and lacks an immediate classical analogue~\cite{Marzari}. Nonetheless, theory has revealed its central role in diverse phenomena, including the anomalous Landau levels~\cite{LL1, LL2, LL3} and superfluid stiffness~\cite{Torma1, Bockrath,Torma2, Bernevig, Torma3} of flat bands, nonlinear Hall effects~\cite{BCdipole, JAhn, QMdipole1, QMdipole2}, excitonic Lamb shifts~\cite{exciton}, and geometric orbital susceptibility~\cite{orbital1, orbital2}. Both papers~\cite{Yang1, Yang2} emphasize that these effects cannot be captured by Berry curvature alone, underscoring the need for experimental access to the quantum metric itself .

Despite its importance, measuring the quantum metric in solids has been fundamentally challenging. Unlike the Berry curvature, which can often be related to dichroic signals or transport coefficients, the quantum metric depends sensitively on momentum derivatives of the Bloch wavefunctions. This requires access to the internal structure---orbital, sublattice, or spin texture---of electronic states with high fidelity, an intrinsically phase- and matrix-element-sensitive measurement.
Photoemission spectroscopy, especially modern synchrotron ARPES, is uniquely positioned here: it measures momentum-resolved spectral weight while allowing systematic control of photon polarization and (in spin-ARPES) spin selection. Two key conceptual moves in the two recent works~\cite{Yang1, Yang2} are as follows: One is to identify measurable geometric quantity that faithfully represents the quantum geometric tensor. The other is to treat polarization and spin dependence of ARPES signal not as nuisance matrix element effects, but as tomographic handles that encode the internal texture of Bloch states. Once that texture is reconstructed, the quantum metric and Berry curvature follow as well-defined functions of momentum.

\begin{figure*}[t!]
\includegraphics[width=17 cm]{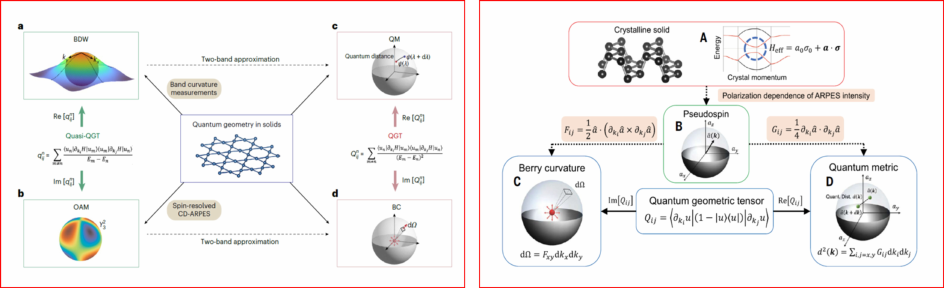}
\caption{
Illustration of the strategies for extracting the quantum geometric tensor from ARPES measurement.
(Left) By introducing quasi-QGT. Adapted from Ref.~\cite{Yang1}. (Right) By measuring pseudospin texture. Adapted from Ref.~\cite{Yang2}.
}
\label{sphere_patch}
\end{figure*}

{\it Toward full QGT metrology via a quasi-QGT bridge in CoSn.}---
Kang et al.~\cite{Yang1} address this challenge by developing a general framework applicable to multiband crystalline solids. Their central conceptual innovation is the introduction of a quasi--quantum geometric tensor, whose real and imaginary parts correspond to experimentally accessible quantities: the band Drude weight and the orbital angular momentum of Bloch electrons. In two-band systems, this quasi-QGT is exactly proportional to the true QGT, while in multiband settings it provides an excellent approximation when bands are energetically isolated.
Experimentally, this framework translates into a two-pronged strategy. The real part of the quasi-QGT is extracted from detailed band-curvature measurements using high-resolution ARPES, while the imaginary part is obtained from circular-dichroism and spin-resolved ARPES, which probe orbital angular momentum and hidden Berry curvature in individual spin channels .

The authors demonstrate this approach in the kagome metal CoSn~\cite{CoSn}, a system hosting Dirac dispersions and flat-band features where quantum geometry is expected to be especially consequential. The promise here is not merely to ``see a number'',but to obtain momentum- and energy-resolved maps of geometric fields in a setting where geometry is believed to control correlated and topological responses.
By reconstructing both the quantum metric and the spin-resolved Berry curvature near the avoided band crossing at the Brillouin-zone center, they reveal the nontrivial geometry underlying the kagome flat bands. The excellent agreement with tight-binding and first-principles calculations confirms that the measured geometric quantities faithfully capture intrinsic Bloch-state properties.
This is an important first-step complemented by the follow up Science paper~\cite{Yang2} where black phosphorus was used for direct extraction of QGT via pseudospin tomography. The quasi-QGT based measurement in CoSn emphasizes generality, aiming to encompass a wide class of materials with more complex internal structure.

{\it Direct quantum-metric tensor tomography in black phosphorus.}---
Kim et al.~\cite{Yang2} report what is essentially the first full-tensor ``map'' of the quantum metric for Bloch electrons in a natural solid.
Kim et al. have addressed this challenge by exploiting a key simplification: when low-energy electronic states are well described by an effective two-band Hamiltonian, the Bloch states can be mapped onto a pseudospin living on the Bloch sphere. In such systems, both the Berry curvature and the quantum metric can be expressed directly in terms of the momentum-space texture of this pseudospin.

In this context, bulk black phosphorus~\cite{BP1,BP2,BP3,BP4} provides an ideal platform. Near the valence-band maximum, its electronic structure is dominated by two well-isolated bands with weak spin--orbit coupling and strong crystalline symmetries~\cite{BPEzawa}. These symmetries constrain the pseudospin to lie on a great circle of the Bloch sphere, allowing it to be parameterized by a single rotation angle. By systematically measuring the linear and circular polarization dependence of ARPES intensity, the authors reconstruct this pseudospin angle throughout momentum space, effectively performing pseudospin tomography. From this reconstructed texture, the full quantum metric tensor---comprising three independent components in two dimensions---is obtained directly. A key result is the pronounced anisotropy of the metric, which is far stronger than the anisotropy of the energy dispersion itself. This vividly illustrates a central message of quantum geometry: band energies alone do not determine geometric properties. The experimental metric maps agree quantitatively with density functional theory calculations, validating both the methodology and the underlying geometric framework.
Importantly, in black phosphorus the Berry curvature vanishes identically due to space--time inversion symmetry~\cite{BP4}. As a result, the quantum metric fully characterizes the QGT, providing a particularly clean demonstration of metric physics in a real solid.

Conceptually, this is powerful for two reasons. First, it demonstrates that all independent components of the quantum metric---not only traces or symmetry-protected combinations---can be accessed experimentally in a bulk solid. Second, it reframes ARPES matrix elements as a feature: the polarization dependence becomes the ``readout channel'' for the underlying Hilbert-space orientation of the Bloch eigenstates~\cite{Pseudo1,Pseudo2,Pseudo3,Pseudo4}. In a field where quantum geometry has often been invoked to explain puzzling responses, this work supplies a missing ingredient: geometry can now be measured first, and only then used to predict or interpret responses.

{\it What these advances unlock.}---
First, geometry-first materials diagnosis became possible. Instead of inferring quantum geometry from downstream responses such as transport coefficients and optical nonlinearities, one can now envision a workflow where QGT maps serve as a primary characterization layer---much like dispersion and Fermi surfaces do today.
Seond, ``geometric mechanisms'' can be quantitatively tested.  A recurring challenge in quantum materials is mechanism ambiguity: does an observed nonlinear Hall signal come from Berry curvature dipoles, disorder, or geometric metric effects? With experimental metric and Berry curvature maps in hand, the geometric part of the theory becomes a falsifiable input rather than a post-hoc explanation.
Third, targeted search for geometry-enabled functionality is feasible. Anomalous flat-band Landau levels, flat-band superconductivity, excitonic structure shifts, and unusual magneto-optical responses have all been argued to depend on quantum geometry. The ability to measure the relevant fields should enable rational screening: find bands with large, anisotropic metric components or sharply structured curvature, then ask which responses track those features.

{\it Open directions and practical challenges.}---
Despite the robustness of the conceptual and technical foundations laid by the two papers~\cite{Yang1, Yang2}, several challenges will shape the next phase.
First, how to go beyond effective two-band descriptions? The quasi-QGT based approach is reliable when two bands under investigation are well-separated from other states in energy and momentum space. Also, black phosphorus benefits from a relatively clean pseudospin reduction near the Fermi level. Extending ``full-tensor'' measurement to strongly entangled multiband manifolds will require robust protocols for disentangling and matrix-element calibration---precisely where the quasi-QGT strategy may become essential.
Second, how to improve resolution, noise, and derivatives? The metric involves momentum derivatives of state textures; experimental extraction will always amplify noise and systematic drift. Community standards---error propagation, consistency checks against sum rules, and cross-validation with first-principles computations---will matter as much as conceptual novelty.
Third, how to connect to many-body renormalization? ARPES measures spectral functions with self-energy effects; quantum geometry is formally defined for eigenstates. A key opportunity is to clarify when measured ``geometry'' should be interpreted as that of effective quasiparticles, and when interaction-driven incoherence demands a generalized (perhaps frequency-dependent) notion of geometry. Finally, how to extend the domain of QGT from momentum space to the hybrid time, position, and momentum space? Generalizing the two ideas used to obtain the momentum-resolved QGT to various experimental techniques with time, position, and momentum resolution will open a new direction to hybrid multi-dimensional geometric responses in solid state platforms.

{\it Outlook.}---
Much as ARPES once transformed band structure from a theoretical construct into a directly observable quantity, the experiments of Kang et al.~\cite{Yang1} and Kim et al.~\cite{Yang2} mark the beginning of an era in which the geometry of Bloch wavefunctions themselves becomes experimentally tangible. Quantum distance, long confined to abstract Hilbert space, has finally entered the laboratory.

In retrospect, it is striking that solid state physics could build an entire topological era largely on Berry curvature~\cite{Niu}---without routinely measuring it as a band-resolved field. The 2025 breakthroughs suggest that quantum geometry is now following the opposite trajectory: measurement is catching up quickly enough to steer theory, not merely confirm it.

If ARPES once transformed band structure from an inferred concept into a photographed reality, quantum-metric and QGT metrology may do the same for the shape of wavefunctions in Hilbert space. The immediate payoff is better mechanistic clarity for quantum geometric responses. The longer-term promise is deeper: a new ``geometric materials science'', where we engineer not only dispersions and symmetries, but also quantum metric and Berry curvatures---and do so with the confidence that these quantities can be measured, mapped, and optimized in the laboratory.

\begin{acknowledgments}
{\it Acknowledgment.|}
B.-J.Y was supported by Samsung Science and Technology Foundation under project no. SSTF-BA2002-06, National Research Foundation of Korea (NRF) funded by the Korean government(MSIT),  grant no. RS-2021-NR060087 and RS-2025-00562579,  Global Research Development-Center (GRDC) Cooperative Hub Program through the NRF funded by the MSIT,  grant no. RS-2023-00258359, Global-LAMP program of  the NRF funded by the Ministry of Education, grant no.  RS-2023-00301976.
\end{acknowledgments}


\providecommand{\newblock}{}

\end{document}